\documentclass[prl,twocolumn]{revtex4}

\usepackage{graphicx}
\usepackage{amsmath}

\begin{document}

\title{Bose--Einstein condensates in traps of time--dependent topology}
\author{Th.~Busch$^{1}$, J.R.~Anglin$^{2}$, and W.H.~Zurek$^{3}$}
\date{\today}

\begin{abstract}
  Superfluid phenomena can be explained in terms of the topologies of
  the order parameter and of the confining vessel. For example,
  currents in a toroidal vessel can be characterized by a discrete and
  conserved quantity, the winding number. In trapped Bose--Einstein
  condensates, the topology of the trap can be characterized by the
  topology of the Thomas--Fermi surface of its N--particle ground
  state. This can be altered during an experiment, so that a toroidal
  trap may deform into a more spherical shape, allowing an initially
  persistent current to decay into singly--quantized vortices. We
  investigate such a procedure numerically, and confirm that the
  Thomas--Fermi prescription for the trap topology gives an accurate
  picture of vortex formation.
\end{abstract}

\affiliation{$^{1}$Institute of Physics and Astronomy, Aarhus University, 
                   Ny Munkegade, DK--8000 Aarhus C, Denmark\\
             $^{2}$Center for Ultracold Atoms, MIT 26--2, 
                   77 Massachusetts Avenue, Cambridge, MA 02139\\
             $^{3}$T--6 (Theoretical Astrophysics), MS B288, 
                   Los Alamos National Laboratory, Los Alamos, 
                   New Mexico 87545}
\maketitle
                 
The intriguing beauty of the concept of topological quantum numbers
has ever since its introduction into physics drawn great attention to
states that are topologically nontrivial \cite{Thouless}. In a system
that has undergone a second order phase transition these numbers
characterize the order parameter field.  For a Bose--condensed system
the order parameter is the phase of the macroscopic wave--function,
and vortex line defects, as well as persistent currents in
multiply--connected vessels, are examples of topologically nontrivial
states that have been well investigated in superfluid $^{4}$He and
$^{3}$He \cite{Donelly}. Indeed, these classic superfluids are
sufficiently strongly interacting systems that the topology of the
order parameter field becomes especially important, because it is
insensitive to the microphysical details for which we lack simple and
reliable theories. Dilute samples of Bose--condensed alkali atoms
\cite{BECs}, however, are weakly interacting, and the numerically
tractable Gross--Pitaevskii (GP) mean field theory describes them with
high quantitative accuracy. They are also amenable to a wide range of
experimental controls, including the time--dependent manipulation of
the inhomogeneous external trapping fields. In this paper we propose
an experimental approach which takes advantage of this capability to
probe the continuous range of physics between vortices and persistent
currents, by adiabatically deforming a trapping potential so that the
topology it imposes on the condensate effectively changes. This causes
an initially uniform current to break up, at some point, into a number
of vortices. The technique thus offers an adiabatic complement to the
stirring experiments on critical velocity in condensates
\cite{Onofrio}.

\begin{figure}[tbp]
  \includegraphics[width=\linewidth,clip=true]{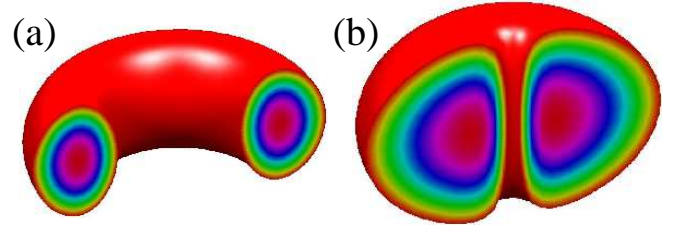}
  \caption{Density of (a) a persistent current in a toroidal trap with $h_t=25$
    and $\protect\sigma^2=5$ (see eq.~(\ref{eq:MexHatPot})) and (b) a
    vortex state in a harmonic trap. For both plots
    $\protect\kappa=1$.}
\label{fig:VPC}
\end{figure}

The topological quantum number of a vortex defect or a persistent
current can be found by writing the wave--function of the superfluid
as the product of a modulus and a phase,
$\Psi=\sqrt{\rho}e^{i\theta}$. The condition of superfluidity means
that the density field $\rho$ is strongly fixed energetically, so that
the remaining low--energy degrees of freedom are those of the current
velocity field, given by the gradient of the phase,
$\vec{v}=(\hbar/m)\vec{\nabla}\theta $, with $m$ being the particle
mass. Because $\theta $ must be single--valued modulo $2\pi $, the
circulation calculated by integrating $\vec{v}$ along an arbitrary
closed contour $C$,
\begin{equation}
\Gamma _{C}=\frac{\hbar}{m}\oint\limits_{C}d\vec{l}\;\vec{\nabla}\theta%
           =\frac{\hbar }{m}\Delta \theta \Big\lvert_{C}\;,
\end{equation}
must be $2\pi (\hbar /m)\kappa $. The integer $\kappa $ will be
unchanged by any continuous deformation of the field $\theta $: it is
a topological invariant known as the \textit{winding number}. If the
entire superfluid sample has no `holes', however, then the contour $C$
can be shrunk to zero size, so that $\kappa $ is changed to zero by a
continuous process, after all. The paradox is resolved by the fact
that the winding number can change if $\rho$ vanishes at any point on
the contour, because there $\theta $ becomes undefined. Hence a simply
connected sample with $\kappa \neq 0$ must contain such topological
defects -- vortex cores. A multiply connected sample can have a
winding number without defects, in which case the current cannot decay
-- unless vortices nucleate at the sample surface, and then drift
across the current. Thus the interplay between topological defects in
the order parameter, and the topology of the confining vessel, gives a
qualitative explanation of superfluidity, which is a necessary
preliminary to any microscopic considerations (of how vortices
nucleate, etc.).

For liquid helium there is a sharp distinction between topological
defects, which are microscopically small, and the topology of the
sample's macroscopic container. For dilute Bose--Einstein condensates,
in contrast, a vortex core may be only a couple of orders of magnitude
smaller than the entire sample. So if we are to use trap topology to
get a simple understanding of how cold atomic currents can be stable,
and how they can decay, then we must be precise about how to
distinguish the `intrinsic topology' of a condensate, due to vortices,
from the topology which is imposed by the trap. Since topology
characterizes closed surfaces, trap topology must be the topology of
an equipotential surface $V(\vec{r})=V_{0}$ . For traps of finite
strength, this topology will in general depend on the choice of
$V_{0}$. For a given number of trapped atoms $N$, however, the natural
candidate is the ground state Thomas--Fermi (TF) surface
$V(\vec{r})=\mu $, where $\mu (N)$ is the condensate chemical
potential. It is a main result of this paper to show that this
prescription does indeed provide a very useful measure of the effect
of the trap on vortex nucleation in circulating condensates. The
distinction between a condensate with a vortex, and a condensate
circulating in a toroidal trap, remains from one point of view only a
matter of degree (see Fig.~\ref{fig:VPC}); but this definition of trap
topology captures the fact that there is a physically important
difference of scale.

\begin{figure}[tbp]
  \includegraphics[width=\linewidth,clip=true]{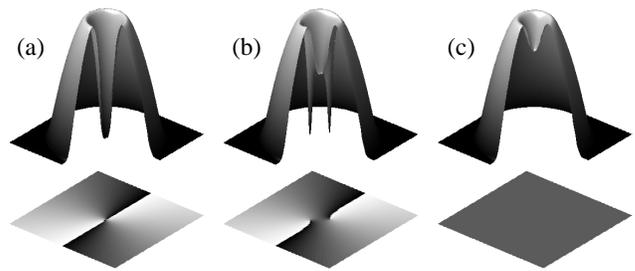}
  \caption{Breaking up of a $\protect\kappa =2$ current, showing the
    respective density (upper row) and phase (lower row)
    distributions.  The height $h_{t}$ is lowered from $(a)$ to $(c)$,
    with $\protect\sigma $ constant, yielding $(a)$ the stable
    $\protect\kappa =2$ current, then $(b)$ a locally stable state
    with two $\protect\kappa =1$ vortices, and finally $(c)$ the
    non--rotating ground state.}
\label{fig:Decay}
\end{figure}

We simulate the evolution of a dilute Bose--Einstein condensate held
initially in a toroidal trap \cite{Bloch73,Yoo98,Geller98}, obtained
by focusing a blue--detuned laser beam into the center of a
rotationally symmetric harmonic trap
\cite{Onofrio,Ketterle98,Sanpera00}. For numerical tractability we
consider the limit of a quasi--two--dimensional trap, and neglect the
third dimension entirely. Assuming the profile of the laser to be
Gaussian, the resulting potential $V_{t}(r)$ can be written as
\begin{equation}
V_{t}(r)=\frac{1}{2}r^{2}+h_{t}e^{-\frac{r^{2}}{\sigma ^{2}}}\;,
\label{eq:MexHatPot}
\end{equation}
where $h_{t}$ and $\sigma $ determine the height and width of the
central peak. The subscript $t$ indicates that the height is to be
modified adiabatically during the evolution \cite{Jila2C}. Using
natural units we scale the length in units of the harmonic oscillator
ground state size $a_{0}=\sqrt{\hbar /m\omega }$ and the energy in
units of the trap frequency $\hbar\omega $. Because the potential is
changed adiabatically, the state of the system at time $t$ is assumed
to be determined by the instantaneous time--independent GP equation
\begin{equation}
 \mu_{t}\psi_{t}=-\frac{1}{2}\Delta_{2}\psi_{t}+V_{t}\psi_{t}
                  +g|\psi_{t}|^{2}\psi_{t}\;,  \label{eq:GPE}
\end{equation}
where the non--linear coupling constant in the quasi--2D trap is given
by $g=4\pi a/a_{0}$, with $a$ representing the atomic $s$--wave
scattering length. We calculate the sequence of instantaneously
stationary states of the adiabatically evolving system by propagating
the time--dependent GP equation in imaginary time \cite{SpStFFT}, and
changing the potential slightly in every time step. We begin with an
initial $\psi _{0}$ with $\kappa >1$, with $h_{0},\sigma _{0}$ such
that this state is (locally) stable. As we slowly lower $h_{t}$, we
find in all cases that at some critical height $h_{c}$ instability
occurs. In Fig.~\ref{fig:Decay} we show a generic example of this
effect, in which a current with $\kappa =2$ breaks up into two
vortices. (The rotational symmetry is broken through numerical noise,
with vortices forming at angles unrelated to the spatial grid.)

\begin{figure}[tbp]
  \includegraphics[width=\linewidth,clip=true]{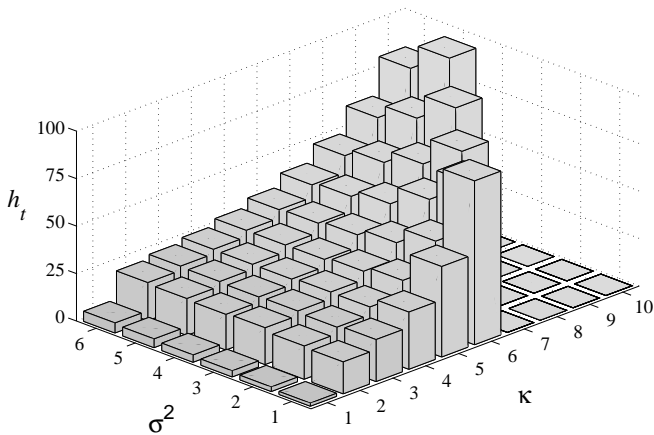}
  \caption{Stability and instability as a function of $\protect\kappa $ and $%
    \protect\sigma ^{2}$. States 'inside' the bars are unstable, $h_c$
    increases with $\protect\kappa $ and decreases with
    $\protect\sigma$ The chemical potential increases from
    $\protect\mu \approx 18$ to $\protect\mu \approx 27$ with
    increasing $\protect\kappa $ and is independent on the scale of
    the plot of $\protect\sigma $. Fields where $h_{t}=0$ (large
    $\protect\kappa $, small $\protect\sigma $) were beyond our
    numerical capabilities because the steepness of the central peak
    necessary to obtain a stationary states could not be resolved
    sufficiently with reasonable spatial grid size. Here
    $g|\protect\psi|^{2}=1000$.}
\label{fig:StabPlot}\centering
\end{figure}

Our goal is to investigate quantitatively the behaviour of this
critical height $h_{c}$ as a function of $\kappa $ and $\sigma $.
Although the imaginary time solution of eq.~(\ref{eq:GPE}) does give
the correct evolution of the system, the vortex cores form at such low
densities that our numerical resolution is insufficient to determine
the precise barrier height at which they first form. We overcome this
limitation by numerically solving the imaginary time Bogoliubov
equations around the solutions of the GP equation, using the ansatz
\begin{equation}
  \Psi_{t}=\psi_{t}+\varphi \;,
\end{equation}
so that the imaginary time Bogoliubov equation reads
\begin{align}
(\mu +\varepsilon )\varphi =& -\frac{1}{2}\left( \partial _{rr}+\frac{1}{r}%
  \partial _{r}+\frac{1}{r^{2}}\partial _{\phi \phi }\right) \varphi
\notag
\label{eq:BOG} \\
& +V(r)\varphi +2U|\psi _{t}|^{2}\varphi +U|\psi _{t}|^{2}\varphi ^{\ast }.
\end{align}
We solve Eq.~(\ref{eq:BOG}) for the lowest eigenvalue $\varepsilon $
by relaxation in imaginary time, using a higher order Runge--Kutta
method. The eigenfunctions $\varphi $ found by this procedure have no
direct physical significance; but the first appearance of a negative
$\varepsilon $ indicates precisely when instability occurs. The nature
of the instability is then determined by the imaginary time GP
evolution as outlined above. We repeat these calculations for a range
of $\kappa $ and $\sigma $; the results are shown in
Fig.~\ref{fig:StabPlot}.

The most salient feature of the results is a `floor' effect: in many
cases, with wider barriers and higher winding numbers, the critical
barrier height is equal to the chemical potential. The critical
barrier height may rise above this `floor', if the barrier is narrow
enough or the winding number small enough. These two basic results
suggest one natural interpretation: the spawning of the vortices is
governed by two factors, one of which is the topology of the ground
state TF surface. We show below that these two factors are
conceptually worth distinguishing, but are not actually physically
independent.

\begin{figure}[tbp]
  \includegraphics[width=\linewidth,clip=true]{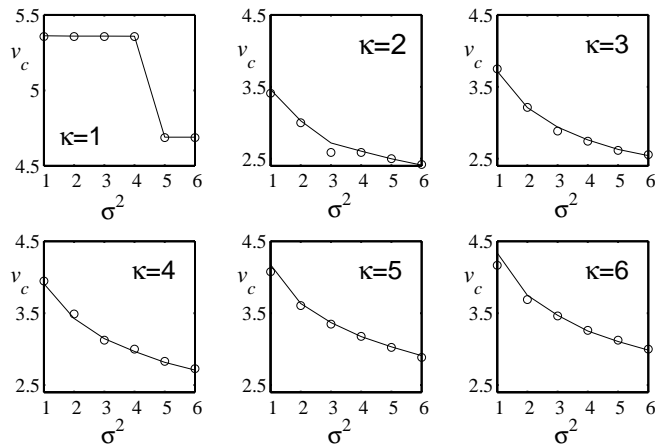}
  \caption{Critical velocity of the current at the inner TF surface as a
    function of the width of the central potential. Circles: numerical
    solutions of eq.~(\ref{eq:BOG}); Full lines: TF approximation as
    described in the text with the proportionality constants
    $A(\protect\kappa)=[2,2.5,2.9,3.3,3.8,3.9]$. Note the different
    scale in the plot for $\protect\kappa =1$.}
\label{fig:CritVel}
\end{figure}

The first factor in vortex formation is instability of surface
excitations \cite{Smith99}. If the velocity of the atoms at the inner
border exceeds the local critical velocity, $v_{c}$, vortices form
while the sample is still multiply connected on the trap scale.
Lowering the central barrier shrinks the inner radius of the toroidal
TF surface, and with fixed circulation this raises the velocity at
this inner radius. If $\kappa $ is high enough, stability may fail
when the inner toroidal radius is so large that the curvature of the
inner TF surface is irrelevant, and we are effectively seeing the
critical velocity for flow past a straight TF surface. In this limit
we can expect to recover the critical velocity for flow perpendicular
to a linear ramp potential (because the vortices will nucleate within
a narrow boundary layer, within which the potential is effectively
linear). From dimensional analysis it then follows that
$v_{c}\propto\left( \hbar F/m^{2}\right) ^{1/3}$ where $F$ is the
derivative of the trapping potential, at the TF surface, in the
direction normal to that surface \cite {Smith99}. A comparison of this
approximation with the results from the numerical solution of
eq.~(\ref{eq:BOG}) is shown in Fig.~\ref{fig:CritVel}.  With
appropriately chosen proportionality constants, which evidently
reflect the role of curvature, the agreement is excellent. Since the
scaling of the critical velocity with the surface force persists quite
well in almost all of our cases, it is evident that the picture of
vortices forming through surface instability is at least qualitatively
valid even in cases where the concept of a surface may well be suspect
(such as those in which the condensate density is very low where the
vortices form).

We can supplement our understanding of vortex formation in this
doubtful regime, and explain the floor effect in
Fig.~\ref{fig:StabPlot}, by invoking trap topology as our second
factor. Lowering $h_{t}$ below the chemical potential abruptly changes
the trap topology, and turns a persistent current of winding number
$|\kappa |>1$ into a multiply quantized vortex. As is well known, a
multiply quantized vortex is unstable against breaking up into unit
vortices, because the energy of a single vortex is
$\propto\kappa^{2}$. Of course, this topological factor is not a
separate mechanism, in addition to the microphysical mechanisms of
instability at supercritical velocity. It should be obvious that as
the inner TF surface disappears, the velocity at that surface becomes
arbitrarily large. But trap topology is a concise summary of
microphysical details, which can quickly identify a whole regime in
which unit vortices will appear.

On the other hand, there can be qualitative features of vortex
behaviour which depend on more specific details of the trapping
potential than its topology. Our results corroborate previous
observations that instability to vortex formation need not mean that
the resulting vortices penetrate the current \cite{Feder99}.
Fig.~\ref{fig:Decay}(b) is not an intermediate stage in a real--time
process, but a local energy minimum for a particular potential height.
It shows that vortices may be stably trapped near the inner TF
surface. From the point of view of superfluidity, this indicates that
a persistent current may be nonlinearly stabilized despite being
linearly unstable. (That is, the linear instability can saturate
nonlinearly without destruction of the current.) From the point of
view of vortex dynamics, this trapping feature would allow study of
interactions between vortices. Since the energy of a vortex scales
with the density of the background fluid, vortices tend to `float'
towards the potential barrier \cite{VortexTrapping}; but the
well--known Coulomb--like repulsion between vortices circulating in the
same direction tends, in the circular geometry, to drive them radially
outwards. A too gentle slope of the central potential, or too many
vortices, allows this repulsion to dominate buoyancy, and the vortices
are expelled through the outer TF surface. As a result the winding
number of the current circulating in the trap is decreased by the
number of exiting vortices.  So one corollary application of our
topology--changing procedure is stabilizing vortices against drifting
out of the sample.

The tunable central barrier can also be useful for creating
circulating currents in the first place. Simulations indicate that
phase imprinting \cite{Ertmer99} methods should be particularly
enhanced, because the problematic core region of the mask becomes
irrelevant. Phase imprinting should also be effective in even less
trivial topologies:~with two Gaussian barriers, and a slightly more
complicated mask, one can make vortex pairs, or vortex--anti--vortex
pairs. And once quantized circulation has been established around the
barriers, they may be moved independently. Finally, the
non--equilibrium phenomenon of spontaneous creation of winding number
in rapid quenches has been predicted to be more readily observable in
toroidal traps \cite{RapidAnglin}, and with a tunable barrier one
could assess this dependence quantitatively.

In summary, therefore, we have proposed and examined a method of
effectively changing the topology of a trap confining a condensate
with a persistent current. We have computed critical parameters for
vortex formation, and identified the effect of trap topology. The
proposed technique would allow the precisely controlled production of
vortices, and may help to shed new light on the general role of
topology in superfluid phenomena.

\begin{acknowledgments}
  We gratefully acknowledge discussion with J.~I.~Cirac, P.~Zoller,
  K.~M\o lmer, and R.~Onofrio.  This work has been supported by the
  European Union under the TMR network No.~ERBFMRX--CT960002, by the
  Austrian FWF, by the Danish research council, and by the American
  NSF through its grant for ITAMP at the Harvard--Smithsonian Center
  for Astrophysics.
\end{acknowledgments}

% --------------------------------------------------------


\begin{thebibliography}{99}
% Topological Quantum Numbers in Nonrelativistic Physics

\bibitem{Thouless}  D.~J.~Thouless, \textsl{Topological Quantum Numbers in
Nonrelativistic Physics} (World Scientific, Singapore, 1997).

% Quantized Vortices in Helium II

\bibitem{Donelly}  R.~Donelly, \textsl{Quantized Vortices in Helium II}
(Cambridge University Press, Cambridge, U.K., 1991).

% BEC's

\bibitem{BECs}  M.~H.~Anderson \textsl{et al.}, Science \textbf{269}, 198
(1995); C.~C.~Bradley \textsl{et al.},Phys.~Rev.~Lett.~\textbf{75}, 1697
(1995); K.~B.~Davis \textsl{et al.}, Phys.~Rev.~Lett.~\textbf{75}, 3969
(1995).

% stirring condensates

\bibitem{Onofrio}  C.~Raman, M.~K\"{o}hl, R.~Onofrio, D.~S.~Durfee,
C.~E.~Kuklewicz, Z.~Hadzibabic, and W.~Ketterle, Phys.~Rev.~Lett.~ \textbf{83%
}, 2502 (1999); R. Onofrio, C. Raman, J. M. Vogels, J. R. Abo--Shaeer, A. P.
Chikkatur and W. Ketterle, Phys. Rev. Lett. \textbf{85}, 2228 (2000). 
%  See also the numerical simulations by 
B. Jackson, J.F. McCann, and C.S. Adams, Phys. Rev. \textbf{A61}, 051603(R)
(2000).

% Superfluidity on a ring

\bibitem{Bloch73}  F.~Bloch, Phys.~Rev.~A \textbf{7}, 2187 (1973).

% Persistent current in a toroidal trap

\bibitem{Yoo98}  J.~Javanainen, S.~M.~Paik and S.~M.~Yoo, Phys.~Rev.~A 
\textbf{58}, 580 (1998).

% % Multiply connected Bose-Einstein-condensed alkali-metal gases:
% % Current carrying states and their decay

\bibitem{Geller98}  E.~J.~Mueller, P.~M.~Goldbart and Y.~Lyanda--Geller,
Phys.~Rev.~A \textbf{57}, R1505 (1998)

\bibitem{Ketterle98} In D.~M.~Stamper--Kurn \textsl{et al.},
  Phys.~Rev.~Lett.~\textbf{81}, 2194 (1998) adiabatic changing of the
  intensity of a laser beam focused into a harmonic trap is reported
  
\bibitem{Sanpera00} In cond--mat/0005136 Martikainen \textsl{et al.}
  suggest that an effective toroidal trap can be realized by rapidly
  shaking a harmonic trap.
  
\bibitem{Jila2C} The early stages of the experiment reported on in
  Matthews \textsl{et al.}, Phys.~Rev.~Lett.~\textsl{83}, 2498 (1999)
  can be seen as a realization of the proposed idea, with the
  non--rotating condensate component providing the central potential.

% Vortices in a Bose-Einstein Condensate
% \bibitem{Cornell99} M.~R.~Matthews, B.~P.~Anderson, P.~C.~Haljan,
%   D.~S.~Hall, C.~E.~Wieman and E.~A.~Cornell,
%   Phys.~Rev.~Lett.~\textbf{83}, 2498 (1999).

% Split-Step Methods for the Solution of the Nonlinear Schr\"odinger Equation
  
\bibitem{SpStFFT} J.~A.~C.~Weideman and B.~M.~Herbst, SIAM
  J.~Numer.~Anal.  \textbf{23}, 485 (1986)

% The Surface of a Bose-Einstein Condensed Atomic Cloud

\bibitem{Smith99}  U.~Al Khawaja, C.~J.~Pethick and H.~Smith, Phys.~Rev.~A~%
  \textbf{60}, 1507 (1999).

\bibitem{Feder99}  D.L. Feder, C.W. Clark, and B.I. Schneider, Phys. Rev. A 
\textbf{61, }011601 (R), (2000).

\bibitem{VortexTrapping} One can see from Fig.~\ref{fig:StabPlot} that
  vortices with $\kappa =1$ are always trapped, unless $h_{t}$ becomes
  so small that the region within which density increases with radius
  becomes smaller than the vortex core.

% Vortex formation in a stirred Bose-Einstein condensate

\bibitem{Dalibard00}  K.~W.~Madison, F.~Chevy, W.~Wohlleben and J.~Dalibard,
Phys.~Rev.~Lett.~\textbf{84}, 806 (2000); J.R. Abo--Shaeer, C.\ Raman, J.M.
Vogels and W. Ketterle, to appear in Science.

% Optical Generation of Vortices in trapped Bose-Einstein Condensates

\bibitem{Ertmer99}  L.~Dobrek, M.~Gajda, M.~Lewenstein, K.~Sengstock, 
G.~Birkl and W.~Ertmer,  Phys.~Rev.~A \textbf{60}, R3381 (1999).

% Vortices in the wake of rapid Bose-Einstein condensation

\bibitem{RapidAnglin}  J.~R.~Anglin and W.~H.~Zurek,  Phys.~Rev.~Lett.~%
\textbf{83}, 1707 (1999).
\end{thebibliography}
\end{document}